 \documentclass{jetpl}
\twocolumn
\title{Topological superfluid $^3$He-B in magnetic field and Ising variable}

\rtitle{Topological superfluid $^3$He-B in magnetic field and Ising variable}

\sodtitle{Topological superfluid $^3$He-B in magnetic field and Ising variable}

\author{
G.E. Volovik
 \/\thanks{
volovik@boojum.hut.fi} 
}

\rauthor{
G.E.Volovik
}

\sodauthor{
Volovik
}

\address{Low Temperature Laboratory, Aalto University, School of Science and
Technology, P.O. Box 15100, FI-00076 AALTO, Finland
\\
 Landau Institute for Theoretical Physics RAS, Kosygina 2,
119334 Moscow, Russia}

\dates{January 16, 2010}{*}

\abstract{
The topological superfluid $^3$He-B provides many examples of the interplay of symmetry and topology. Here we consider the effect of magnetic field on topological properties of $^3$He-B.  Magnetic field violates the time reversal symmetry. As a result, the topological invariant supported by this symmetry ceases to exist; and thus the gapless fermions on the surface of  $^3$He-B are not protected any more by topology: they become fully gapped. Nevertheless, if perturbation of symmetry is small, the surface fermions remain relativistic with mass proportional to symmetry violating perturbation -- magnetic field.  The  $^3$He-B symmetry gives rise to the Ising variable $I=\pm 1$, which emerges in magnetic field  and which characterizes the states of the surface of  $^3$He-B. This variable also determines the sign of the mass term of surface fermions and the topological invariant describing their effective Hamiltonian. The line on the surface, which separates the surface domains with different $I$,  contains $1+1$ gapless fermions, which are protected by combined action of symmetry and topology. 
}

\begin{document}

\maketitle

\section{Introduction}

Recently topological insulators, semimetals, superconductors, superfluids and other topologically nontrivial gapless and gapped phases of matter have attracted a lot of attention, see e.g. \cite{Fu2007,Moore2007,Schnyder2009a,Kitaev2009}. 
Superfluid $^3$He-B  represents the fully gapped topological superfluids with time reversal symmetry \cite{Schnyder2009a,Kitaev2009,Volovik2009b}. 
This $3+1$ system contains $2+1$ gapless Andreev bound states on the surface of the liquid or at the interfaces between the bulk states with different topological invariants; these fermion zero modes resemble $2+1$ Majorana fermions in relativistic quantum field theories \cite{SalomaaVolovik1988,ChungZhang2009,Volovik2009}. Andreev bound states on the surface of $^3$He-B were discussed theoretically (see e.g. \cite{Kopnin1991,Nagai2008,Nagato2009}) and probed experimentally 
\cite{CCGMP,Davis2008,Murakawa2009c}.  
However, the Majorana signature of these states has not yet been observed.
One of the possible tools for observation of the ``relativistic'' spectrum of the bound states  is NMR, which requires an external magnetic field. The influence of magnetic field on  Majorana fermions has been recently discussed in \cite{Nagato2009}. Here we consider the effect of magnetic field in more detail.

The topological invariant for  $^3$He-B which gives rise to the $2+1$ gapless fermion zero modes on its surface is supported by the time reversal symmetry \cite{Volovik2009b}. Magnetic field destroys this symmetry, as a result the Andreev bound states acquire gap (mass) proportional to the magnitude of magnetic field. The resulting massive $2+1$ Andreev-Majorana fermions also have nontrivial topology, described by the nonzero topological invariant in $2+1$ momentum space. The value of this invariant  is determined by the Ising type variable $I$, which characterizes the surface of $^3$He-B in the presence of magnetic field. The surface of $^3$He-B can be in two states, with $I=+1$ and $I=-1$, and both states of the surface have been recently probed by NMR experiments with $^3$He-B confined in a single micron-size slab  \cite{Levitin2010}. 
These experiments also detected the coexistence of the domains with different $I$ on the surface of $^3$He-B; these domains are separated by the one-dimemsional domain walls (lines)  on the surface. 
Since the domains are characterized by different values of the topological invariant, the boundary line between the domains contains $1+1$ gapless fermions with ``relativistic'' spectrum.

\section{Distortion of order parameter}

Magnetic field destroys the rotational symmetry of $^3$He-B, leading to the anisotropic gap. The gap anisotropy is also induced by the walls. 
The order  parameter in  $^3$He-B is $3\times 3$ matrix $A_{\alpha i}$  \cite{VollhardtWolfle}. It transforms as vector under spin rotation (the Greek index), and as vector under orbital rotation (the Latin index).  For the   distorted  $^3$He-B it has the form
\begin{equation}
A_{\alpha i}=R_{\alpha i} A^{(0)}_{ij} ~~,~~A^{(0)}_{ij}=\left(
\begin{matrix}
\Delta_\parallel &0&0\cr
0&\Delta_\parallel &0\cr
0&0&\Delta_\perp \cr
\end{matrix} 
\right)~~,~~  q=\frac{\Delta_\perp}{\Delta_\parallel}\,,
\label{OrderParameter}
\end{equation}
where  $q$ is the distortion parameter  with $q=1$ in the 
non-deformed B-phase and $q=0$ in the planar state;  $\Delta_\perp$ is the gap  for quasiparticles propagating  in direction along the normal to the wall; $\Delta_\parallel$  is the gap for quasiparticles propagating in directions parallel to the wall;  $R_{\alpha i}$ is rotation matrix, which connects spin and orbital spaces. As said, the distortion of the gap is caused both by magnetic field and by boundaries. Let us introduce the unit  vector $\hat{\bf h}={\bf H}/H$ in direction of magnetic field, and the unit vector
$\hat l_i=\hat h_\alpha R_{\alpha i}$ in the orbital space. The orbital vector  $\hat{\bf l}$ shows the direction of the orbital angular momentum of Cooper pairs (see e.g. 
\cite{SalomaaVolovik1987,Dmitriev1999}) and  plays the role of the anisotropy axis for the uniaxial deformation of the gap.  

\section{Ising variable probed by NMR}

Let us consider NMR experiments with superfluid $^3$He-B confined in a single micron size slab  \cite{Levitin2010}, where
the field is normal to the plates: ${\bf H}=H\hat{\bf z}$. 
Typical fields used for NMR exceed the dipole field (spin-orbit coupling). As a result the boundary conditions for $\hat{\bf l}$-vector is that $\hat{\bf l}$ is normal to the walls, $\hat{\bf l}\times \hat{\bf z} =0$
\cite{BrinkmanCross1978}. This boundary condition ensures that the anisotropy axis  $\hat{\bf z}$ caused by the walls and anisotropy axis $\hat{\bf l}$ caused by magnetic field  coincide. 

In a given geometry with the field being normal to the plates, the projection of the $\hat{\bf l}$-vector on the magnetic field direction \begin{equation}
I=\hat{\bf l}\cdot \hat{\bf h}\,,
\label{Ising}
\end{equation}
represents the Ising variable.
Thus is because only two orientations of $\hat{\bf l}$-vector with respect to manetic field are allowed by boundary conditions:  $I  =+1$, and    $I= -1$.  
The degeneracy  is lifted by spin-orbit interaction. The latter is small compared with all other energies but is important for the measured NMR frequency shift from the Larmor value $\omega_L$, which is just caused by spin-orbit interaction. The measurements of the frequency shift allows to determine experimentally the value of the Ising spin $I$ on the surface.

For the distorted $^3$He-B, the corresponding frequency shifts can be found from equations (4.5) and (4.10) of Ref.~\cite{BunkovVolovik1993}. The frequency shift is positive when $\hat{\bf l}$-vector is parallel to the field, i.e. for $I=+1$, 
and negative  for $\hat{\bf l}$-vector antiparallel to field, i.e. for $I =-1$. The ratio of the frequency shifts  is   
\begin{equation}
\frac {\omega(I=+1)-\omega_L}{\omega_L- \omega(I =-1)}=\frac{1-q^2}{1+2q^2}\,.
\label{ShiftsRatio}
\end{equation}
Both positive and negative frequency shifts have been observed in  \cite{Levitin2010}. From Fig.~1 of  \cite{Levitin2010} and Eq.\eqref{ShiftsRatio} it follows that in the  geometry  of experiment, the distortion parameter $q$ at  
temperature $T=0.54T_c$ is about $q\sim 0.6-0.7$. From this figure it also follows that in cooling one obtains three different states:   the  state  with   $I =+1$ characterized by small positive frequency shift;  the state 
with  $I=-1$ characterized by large negative frequency shift;  and the state with coexisting domains  of 
both orientations of the Ising variable, which is characterized by two peaks in the NMR spectrum. The latter state contains domain wall(s) -- the  line(s) on the surface  separating surface domains with opposite orientations of $I$.

Let us also for completeness consider the field parallel to the plates. In this case the $\hat{\bf l}$-vector is  perpendicular to the field and $I=\hat{\bf l}\cdot \hat{\bf h} =0$.  The frequency shift is positive, $\omega(I =0)-\omega_L=-(\omega(I =-1)-\omega_L)>0$.

We have seen that the orientation of the magnetic field with respect to the Ising spin is crucial for the NMR experiments. Let us consider the effect of orientation of the magnetic field on the spectrum of Majorana fermions.

\section{Majorana fermions in distorted B-phase}

Fermions in magnetic field are described by the following Hamiltonian:
\begin{eqnarray}
{\cal H}=\frac{k^2-k_F^2}{2m^*} \tau_3-\frac{1}{2}\gamma {\bf H}\cdot {\mbox{\boldmath$\sigma$}} +
\nonumber
\\
   \tau_1 \left(\Delta_\parallel(z)\tilde \sigma_x  \frac{k_x}{k_F}+ \Delta_\parallel(z)\tilde\sigma_y  \frac{k_y}{k_F} +\Delta_\perp(z)\tilde\sigma_z \frac{k_z}{k_F}  \right)\,,
\label{eq:Hamiltonian}
\end{eqnarray}
where $\tau_i$ are Pauli matrices of Bogolyubov-Nambu spin;   $\sigma_\alpha$   are Pauli matrices of  $^3$He nuclear spin;  $\gamma$ is the gyromagnetic ratio of the $^3$He atom; and finally 
$\tilde \sigma_i=\sigma_\alpha R_{\alpha i}$ are the Pauli matrices obtained from matrices ${\mbox{\boldmath$\sigma$}}$ by rotation. 
In the absence of magnetic field there is a topological invariant for $^3$He-B \cite{Volovik2009b}:
\begin{equation}
 N = {e_{ijk}\over{24\pi^2}} ~
{\bf tr}\left[  \tau_2 \int    d^3k 
~{\cal H}^{-1}\partial_{k_i} {\cal H}
{\cal H}^{-1}\partial_{k_j} {\cal H} {\cal H}^{-1}\partial_{k_k}  {\cal H}\right].
\label{N3+0}
\end{equation}

At ${\bf H}=0$, when the time reversal symmetry is preserved, the integral \eqref{N3+0} is invariant under deformations because the matrix $ \tau_2$ anti-commutes with Hamiltonian ${\cal H}$.  The nonzero value of this invariant supports gapless fermions  at the interface between bulk states with different values of $N$ \cite{SalomaaVolovik1988} and on the surface of $^3$He-B \cite{Volovik2009}. 
These fermions contribute to the ground-state spin supercurrent  \cite{SalomaaVolovik1988}.

In the presence of magnetic field, the integral \eqref{N3+0} does not represent the topological invariant because the Pauli term does not anti-commute with $ \tau_2$, and the gaplessness of the Andreev bound states is not guaranteed. The Pauli term can be expressed in terms of the rotated spin $\tilde{\mbox{\boldmath$\sigma$}}$:
\begin{equation}
{\bf H}\cdot {\mbox{\boldmath$\sigma$}} =H_\alpha\sigma_\alpha=H_\alpha R_{\alpha i}\tilde\sigma_i=H ~\hat{\bf l}\cdot \tilde{\mbox{\boldmath$\sigma$}}=(\hat{\bf l} \cdot \hat{\bf z}) H\tilde \sigma_z  \,.
\label{Identity}
\end{equation}
Equation \eqref{Identity} demonstrates that  the Hamiltonian \eqref{eq:Hamiltonian}  does not depend on the orientation of magnetic field. It depends only on the Ising variable $\hat{\bf l} \cdot \hat{\bf z}=\pm 1$, which describes two possible orientations of   the vector $\hat{\bf l}$ on the surface. This is because in large fields used in experiments,  the spin-orbit interaction can be neglected and the boundary conditions for  the vector $\hat{\bf l}$ on the surface are fully determined by the orbital effects, which are invariant under rotation of the magnetic field together with the spin subsystem. 

Let us, however, consider the orientation of magnetic field along the axis $z$. This allows us to express the Hamiltonian in terms of the variable $I=\hat{\bf l} \cdot \hat{\bf h}$ in \eqref{Ising}, which is regulated in NMR experiments in Ref.~ \cite{Levitin2010} where the orientation of the field normal to the plates is used:
\begin{eqnarray}
{\cal H}=\frac{k^2-k_F^2}{2m^*} \tau_3 - \frac{1}{2}\gamma H I \tilde \sigma_z +
\nonumber
\\
   \tau_1 \left(\Delta_\parallel(z)\tilde \sigma_x  \frac{k_x}{k_F}+ \Delta_\parallel(z)\tilde\sigma_y  \frac{k_y}{k_F} +\Delta_\perp(z)\tilde\sigma_z \frac{k_z}{k_F}  \right)\,.
\label{eq:Hamiltonian2}
\end{eqnarray}

The second order secular equation (10) in \cite{Volovik2009}  produces the following $2\times 2$ Hamiltonian for fermion zero modes:  
\begin{equation}
\left(
\begin{matrix}
 H^{(1)}_{++}&H^{(1)}_{+-}\cr
H^{(1)}_{-+}&H^{(1)}_{--}\cr
\end{matrix} 
\right) 
=c \left(
\begin{matrix}
- \frac{1}{2}\gamma H I &k_y+ik_x\cr
k_y-ik_x& \frac{1}{2}\gamma H I \cr
\end{matrix} 
\right) 
 \,,
\label{eq:Effective}
\end{equation}
which can be written as
 \begin{equation}
{\cal H}_{\rm zm} = c \hat{\bf z} \cdot(\tilde{\mbox{\boldmath$\sigma$}} \times {\bf k}) - \frac{\gamma H I}{2} \hat{\bf z} \cdot\tilde{\mbox{\boldmath$\sigma$}}~.
\label{eq:ModesH}
\end{equation} 
This the so-called helical (see e.g. \cite{Zang2010} and references therein) 
Hamiltonian produces the  relativistic spectrum of massive particles:
\begin{equation}
E^2=c^2k^2 +M^2 \,,
\label{eq:MassiveParticle}
\end{equation}
where the  `speed of light'  \cite{Volovik2009}
 \begin{equation}
c=\frac{ \int_0^\infty dz \frac{\Delta_\parallel(z)}{k_F} \exp\left(- \frac{2}{v_F} \int_0^z dz' \Delta_\perp(z')\right)}
{ \int_{0}^\infty dz   \exp\left(- \frac{2}{v_F} \int_0^z dz' \Delta_\perp(z')\right)} 
 \,.
\label{eq:SpeedOfLight}
\end{equation}
The mass of the surface fermions
\begin{equation}
 M=\frac{1}{2}\gamma H \,,
\label{eq:Mass}
\end{equation}
appears due to violation of the time reversal symmetry by magnetic field.
The mass \eqref{eq:Mass} was obtained in Ref. \cite{Nagato2009} only for the case of magnetic field normal to the wall. However, as distinct from Ref. \cite{Nagato2009}, in our case the equation \eqref{eq:Mass} is valid for any orientation of magnetic field.  This is because we consider relatively large magnetic fields used in NMR experiments, when the spin-orbit interaction is relatively small:  though  the spin-orbit interaction is responsible for the frequency shift, it does not influence the orientation of  the vector $\hat{\bf l}$ on the surface. As a result the Pauli term \eqref{Identity} expressed via the rotated spin  does not depend on orientation of magnetic field.  

This is contrary to the effect of magnetic field discussed in Refs.~\cite{ChungZhang2009,Nagato2009}, where the magnetic field is assumed to be so small that the boundary conditions are solely determined by spin-orbit interaction. This leads to  anisotropy of magnetic susceptibility \cite{ChungZhang2009,Nagato2009}, which however disappears in large fields.
Note also that in our case the spin  which enters the Hamiltonian \eqref{eq:ModesH} for Majorana fermions, is obtained from the real spin by rotations provided by the order parameter matrix $R_{\alpha i}$. This means that in large fields, when the spin-orbit interaction can be neglected, the quantization axis for the real spin of Majorana fermions is not fixed.

The effective Hamiltonian \eqref{eq:ModesH} has nontrivial topology described by the topological invariant
\begin{equation}
 N_{2+1} =\frac{1}{4\pi^2} ~
{\bf tr}\left[  \int    d^2k d\omega
~G\partial_{k_x} G^{-1}
G\partial_{k_y} G^{-1}G\partial_{\omega}  G^{-1}\right],
\label{N2+1}
\end{equation}
where $G=1/(i\omega - {\cal H}_{\rm zm})$ is the Green's function of fermion zero modes.
The invariant \eqref{N2+1} was first introduced in relativistic $2+1$ theories 
\cite{IshikawaMatsuyama1986,IshikawaMatsuyama1987,Matsuyama1987}. It is responsible for quantization of Hall conductivity induced by topological structure of energy-momentum space, the so-called  intrinsic quantum Hall effect. It was  later independently introduced for the film of superfluid $^3$He-A in condensed matter 
\cite{Volovik1988,VolovikYakovenko1989}. Under experimental conditions in  Ref.~\cite{Levitin2010}, the value of this invariant  is determined by the Ising variable $I$ in \eqref{Ising}:
\begin{equation}
 N_{2+1} =\frac{I}{2} \,.
\label{N2+1I}
\end{equation}
 NMR experiments in Ref.~\cite{Levitin2010} detected coexistence of  the surface domains with different $I$, which are thus separated by the one-dimensional domain wall on the surface. Since the domains have different values of topological invariant $N_{2+1}$, then according to the index theorem (see e.g. \cite{Volovik2003}), the line separating two domains contains gapless $1+1$ fermion zero modes.

\section{Discussion}

The symmetry of superfluid $^3$He-B is important for its topological properties. Magnetic field violates isotropy of  $^3$He-B:  the gap become anisotropic with $\Delta_\parallel - \Delta_\perp \propto H^2$.
This distortion of the gap does not change the topological invariant introduced in \cite{Volovik2009b}, and thus cannot destroy the massless relativistic fermions on the surface of $^3$He-B. However, the violation of the other symmetry  of  $^3$He-B by magnetic field -- 
 the time reversal symmetry -- is crucial. The topological invariant supported by this symmetry ceases to exist and thus the gapless fermions on the surface of  $^3$He-B are not protected any more by topology: they become fully gapped. Nevertheless, if perturbation of symmetry is small, the surface fermions remain relativistic with mass proportional to symmetry violating perturbation -- magnetic field.  
 
 This is not the whole story. The spontaneously broken relative spin-orbit symmetry of $^3$He-B \cite{Leggett1973} -- symmetry under relative rotations of spin subsystem with the respect to the orbital one \cite{footnote} --  gives rise to the vector $\hat{\bf l}$ of orbital  anisotropy in the presence of magnetic field. This in turn leads to Ising variable $I=\pm 1$ which characterizes the surface of  $^3$He-B. The Ising variable determines the sign of the mass term of the surface fermions and the topological invariant \eqref{N2+1I} describing their effective Hamiltonian \eqref{eq:ModesH}. The line on the surface, which separates domains with different $I$,  contains gapless fermions, which are   protected by combined action of symmetry and topology.  So,  the superfluid $^3$He-B provides many examples of the interplay of symmetry and topology.

The other examples of the interplay of symmetry and topology in topological matter and in vacua of relativistic quantum fields can be found in \cite{Volovik2003,Volovik2007,Volovik2010}. Some topological invariants exists only due to symmetry, while some values of the topological invariant are not compatible with symmetry.
It is right time for the general classification of topological matter in terms of its symmetry and 
topology. The attempt of such classification was made in \cite{Fu2007,Moore2007,Schnyder2009a,Kitaev2009}. However,
only non-interacting systems were considered, and not all the symmetries were exploited. Examples are again provided by the superfluid phases of  $^3$He, where the symmetry is enhanced due to relative smallness of the spin-orbit interactions. One task should be to consider  symmetry classes, including crystal symmetry  classes, magnetic classes, superconductivity classes, etc., and to find out  
what are the topological classes of Green's function which are allowed within a given symmetry class. Then one should find out what happens when the symmetry is smoothly violated, etc.
The Green's function matrices with spin and band indices must be used for this classification, 
since they take into account interaction. In this way  we may finally obtain the full classification of  
topological matter, including insulators, superconductors, magnets, liquids, and vacua of relativistic quantum fields. 

The problem of the interplay of topology and symmetry exists not only for the momentum-space topology, but  also for the  real-space topology, which considers classification of topological defects and textures in systems with broken symmetry. For example, some values of the winding number describing the topological defect are not compatible with symmetry of this defect, which connects for example the points ${\bf r}$ and $-{\bf r}$. This happens with monopoles \cite{WilkinsonGoldhaber1977};  disclinations -- topological defects in liquid crystals \cite{Balinskii1984}; vortices in superfluids \cite{SalomaaVolovik1987};   cosmic strings; etc.  One of the tasks there is the  classification of topological objects in terms of their symmetry. Similarly we need the symmetry classification of topological objects in momentum space, where the symmetry connects different points in momentum space, say ${\bf k}$ and $-{\bf k}$ \cite{Fu2007}.

The special role  in classification of topological  systems is played by dimensional reduction. See for example  Chapters 11 and 21 in \cite{Volovik2003} where the topological invariants for the 2D fully gapped systems are obtained by dimensional reduction of
the topological invariants for the 3D gapless system; the dimensional reduction has been recently discussed in \cite{Ryu2009}.  
The dimensional reduction allows us to use for classification of the gapped systems  the scheme, which was suggested by Horava for the classification of the topologically nontrivial nodes in spectrum \cite{Horava2005}. But this also must be supplemented by the symmetry analysis, and also by analysis of the types of the emergent symmetries which necessarily appear within some topological classes.

 I owe special thanks to  John Saunders for discussion on the experiments \cite{Levitin2010} and am grateful to Shinsei Ryu and Liang Fu for discussion on symmetry and topology. This work is supported in part  by the Academy of Finland, Centers of Excellence Program 2006--2011,
and the Khalatnikov--Starobinsky leading scientific school (grant 4899.2008.2).


\end{document}